\documentclass[draft]{agujournal2019}
\usepackage{url} 
\usepackage{lineno}
\usepackage[inline]{trackchanges} 
\usepackage{soul}
\draftfalse

\journalname{JGR: Space Physics}

\begin{document}

%
%


\title{Global asymmetry in $\Delta$X variations during the 06 April 2000 geomagnetic storm: Relative roles of IMF $B_z$ and $B_y$}

%
%




\authors{Sumanjit Chakraborty and D. Chakrabarty}


\affiliation{}{Space and Atmospheric Sciences Division, Physical Research Laboratory, Ahmedabad 380009, Gujarat, India}





\correspondingauthor{Sumanjit Chakraborty}{sumanjit11@gmail.com}




\begin{keypoints}

\item First observations of asymmetry in $\Delta$X variations between two pairs of nearly antipodal locations over low latitudes.
\item Increase in the ratio of IMF $B_y$ to $B_z$ is suggested to cause differences of $\Delta$X variations in one pair with respect to the other.
\item Distortions in DP2 cells, rotation of electrodynamical divider and associated throat flows are believed to determine the asymmetry observed.

\end{keypoints}

%
%

%
%


\begin{abstract}

This investigation is directed to understand the asymmetry in $\Delta$X variations caused due to the relative roles played by IMF $B_z$ and IMF $B_y$ in a particular interval (22:22 - 22:55 UT), during the main phase of a strong geomagnetic storm event of April 06, 2000 ($A_p$ = 236). Two pairs of antipodal stations, being part of the SuperMAG network, are considered here. Ionospheric convection maps from SuperDARN network are used to understand spatio-temporal evolution of the DP2 ionospheric convection patterns over high-latitudes. The two-dimensional maps of equivalent currents are used to show signatures of global DP2 currents associated with the interplay effect between the two IMF components. Observations show increases in the difference in $\Delta$X variations between nearly antipodal stations from the Japanese-European/African sector with respect to the same between the nearly antipodal stations from the Pacific/American-Indian sector. This asymmetry is observed during the period when the absolute magnitude of IMF $B_y$ is larger than that of IMF $B_z$ resulting in a significant and conspicuous enhancement in IMF $|By/Bz|$. It is suggested that the distortions in DP2 cells and associated rotation of electrodynamic day-night divider, bring one pair of stations under the same DP2 cell and one station of the other pair under a different DP2 cell and throat flow region leading to the asymmetry in $\Delta$X variations between the antipodal stations. Therefore, the work highlights the importance of the interplay between IMF $B_z$ and IMF $B_y$ in determining the ionospheric impact over low latitudes during strong geomagnetic conditions.

\end{abstract}


%
%

%


%
%
%
%


\section{Introduction}

The interaction of the Earth's Magnetosphere-Ionosphere (MI) system with the continuously emanating solar wind carrying the frozen-in Interplanetary Magnetic Field (IMF) forms the basis of solar-terrestrial coupling and related physical mechanisms. The solar wind energy gets transferred to the magnetosphere by the process of magnetic reconnection. The amount of this energy being transferred depends on the orientation of the north-south component ($B_z$) of the IMF \cite{sc:1}. The coupling between the solar wind and the MI system is strongest, setting conditions favorable for the occurrence of strong geomagnetic storms: Dst$<=$-100 nT \cite{sc:34}, when the IMF $B_z$ orientation becomes completely southward and stays in this configuration for a sufficient interval of time (at least 3 hours) \cite{sc:2,sc:42,sc:3,sc:4}. Due to this interaction, in addition to changes over the polar and high-latitude ionosphere, the quiet-time electrodynamics of the ionosphere over the low-to-equatorial latitudes gets perturbed, via the process of Prompt Penetration of interplanetary Electric Field (PPEF), causing performance degradation of space-based global navigation satellite systems like the GPS over these geosensitive locations (see \citeA{sc:41,sc:49} and references therein). 

Large-scale convection patterns over the polar and the high-latitudes, as a function of the IMF, have been portrayed by researchers (\citeA{sc:28,sc:20} and references therein) utilizing ionospheric measurements. Furthermore, during the 1980s, the 1990s, and the beginning of the 2000s, the IMF dependencies of global convection have been addressed and consolidated by several researchers \cite{sc:29,sc:30,sc:31,sc:21,sc:32,sc:33}. Their studies involved less direct measurements utilizing ground-based magnetometers and the development of statistical models using High-Frequency (HF) radar networks that are ground-based. It is well known that the $B_y$ component of IMF plays a crucial role in the rotation of the two-cell or DP2 plasma convection patterns produced by the $B_z$ component of IMF in the polar and high-latitude regions of the ionosphere (\citeA{sc:5} and references therein). Several other studies \cite{sc:12,sc:13,sc:14,sc:15,sc:16,sc:17,sc:18,sc:19} show that the reconnection geometry in the dayside is affected by IMF $B_y$ while the dawn-dusk asymmetries, of the magnetospheric convection, are caused due to the stress exerted by the $B_y$ component of IMF on the open magnetic field lines in addition to the modulation of the distribution of Field-Aligned-Currents (FACs), in the dayside, by IMF $B_y$. Additionally, under the influence of IMF $B_y$, modulation in the spatial and amplitude variations of the ionospheric electric field is observed \cite{sc:20,sc:21}. Under the influence of varying IMF $B_y$, few studies \cite{sc:35,sc:36,sc:37}, utilizing simulations and observations, show changes in the thermospheric densities, high-latitude hemispheric asymmetries, and wind patterns.

Several studies \cite{sc:44,sc:45,sc:46,sc:47} have well established the crucial role played by IMF $B_z$ in terms of the PPEF from the high- to the equatorial and the low-latitude ionosphere. During geomagnetically active periods, the study by \cite{sc:48} shows PPEF to have an eastward polarity up to 22:00 LT and become westward thereafter. However, as the overshielding electric field perturbation has an opposite polarity (westward/eastward during daytime/nighttime) over the equatorial ionosphere \cite{sc:50,sc:47}, opposite polarities of PPEF are expected over these regions during daytime and nighttime is expected, being consistent with the curl-free condition of the ionospheric electric field. This theory has been explained thoroughly in the studies by \cite{sc:39,sc:40} where they observe opposite polarities of the PPEF over nearly antipodal locations, one at Jicamarca in the Peruvian sector and the other at Thumba in the Indian sector. Due to such dominant effects of IMF $B_z$, it becomes difficult to properly understand the effect of IMF $B_y$ over these regions. Studies by \cite{sc:38,sc:22,sc:43} show modifications in the PPEF polarity during the geomagnetic storm period, the same polarity of PPEF over nearly two antipodal locations: Jicamarca and Thumba, and asymmetry in the ring current respectively as a result of the role played by the $B_y$ component of IMF. In the study by \cite{sc:22}, they show the evolution of the $\Delta$X component (see Figure 3 of that paper) of magnetic field (obtained from SuperMAG) from complete anti-correlation to becoming strongly correlated as one goes from high-latitude to low-latitude antipodal locations, due to varying orientation of IMF $B_y$, under southward IMF $B_z$, which thereby caused alterations of the equipotential contour distribution over the low-latitude ionosphere, thus bringing the two nearly antipodal locations under the same convection cell.

Although there are several studies present in the literature that addresses the dominant effect of IMF $B_z$ over the low-latitude regions during the geomagnetic storm period, there are very few studies that address the relative roles played by both IMF $B_z$ and IMF $B_y$ to affect the magnetic field perturbations over nearly antipodal locations of these regions. The present study, for the first time, investigates how the variation of the $\Delta$X component of the magnetic field would be affected under the combined effects of IMF $B_z$ and IMF $B_y$ in the main phase of a strong geomagnetic event. For this investigation, the strong geomagnetic storm of April 06, 2000, has been taken up as a case study. Observations of $\Delta$X variations, from SuperMAG, for two pairs of nearly antipodal locations (details of which are given in Table \ref{sc00}) distributed across the globe have been utilized for this study. The motivation of this work comes from the fact that earlier studies have showed similar variations of magnetic field over the antipodal locations due to the role of IMF $B_y$ under southward IMF $B_z$, however we wanted to investigate that whether the same would be true for any choice of pairs of antipodal stations under the effects of interplay between the two IMF components. The novelty of this study is in bringing forward the observation of an asymmetry in the magnetic field variations, under geomagnetically disturbed conditions, due to the relative roles played by both IMF $B_y$ and IMF $B_z$. This study is crucial for understanding the changes in global electrodynamics during strong space weather events and would be vital for developing a reliable space weather forecast system.   

The manuscript is divided as follows: Section 2 presents the descriptions of the locations of observation and the data used for the analyses. Section 3 describes the results obtained. Section 4 describes the discussions of these results and observations. Finally, Section 5 presents the summary of the work.

\section{Data and Observation Sites}

In the present study, the locations of the four stations from where the $\Delta$X component of the magnetic field are obtained (post nighttime base level subtraction) for analysis along with the corresponding Magnetic Latitude (MLAT) and the Local Time (LT) with respect to the Universal Time (UT) are given in Table \ref{sc00}. 
The SuperMAG \cite{sc:25,sc:26}, is a global collaboration of national agencies and organizations, that present;y operates more than 300 ground-based magnetometers. They provide ground magnetic field variations from all the available stations in a common coordinate system, with a common baseline removal approach and same time resolution. It utilizes three-dimensional vector measurements of the magnetic field from these ground-based magnetometers. SuperMAG resamples raw data into one-minute temporal resolution and converts all the units into nT. This data is then rotated into a local magnetic field coordinate system and the baseline is subtracted by an automated technique. These magnetic data are available at http://supermag.jhuapl.edu/. 

\begin{table}[ht]
\caption{Locations of the magnetometer stations (from SuperMAG) used for the analysis along with the corresponding MLAT and LT with respect to the UT}
\centering
\begin{tabular}{lcccc}
\hline
Station Name & Station Code & Sector & MLAT (deg) & LT (h)  \\
\hline
Chichijima & CBI  & Japanese         & 20.00 &  UT+09.49  \\
Alibag     & ABG  & Indian           & 12.05 &  UT+04.86  \\
Guimar     & GUI  & European/African & 14.33 &  UT-01.09  \\
Ewa Beach  & EWA  & Pacific/American & 21.44 &  UT-10.53 \\
\hline
\label{sc00}
\end{tabular}
\end{table}

Furthermore, the ionospheric convection maps used in the present work are obtained from the Super Dual Auroral Radar Network \cite{sc:27} or SuperDARN which consists of more than 30 low-power high-frequency radars. They observe Earth's upper atmosphere from the mid-latitudes to the polar regions. These radars continuously operate to provide valuable information on the near-earth space environment by observing the plasma dynamics occurring in the ionosphere. These convection maps are available at http://vt.superdarn.org/.

Additionally, for observations of the interplanetary, auroral and geomagnetic conditions, the high-resolution (1 minute) IMF $B_z$ and IMF $B_y$ (nT), the solar wind flow pressure (nPa) along with the Polar Cap index (PCN), the Auroral Upper and Lower (AU and AL, nT) and the SYM-H and the ASYM-H (nT) data are all obtained from the NASA OMNIWEB portal at 
https://omniweb.gsfc.nasa.gov/form/omni\_min.html. The SuperMAG AU and AL indices are obtained from http://supermag.jhuapl.edu/.

\section{Geomagnetic storm of April 06, 2000: Results}

\begin{figure}[ht]
\centering
\includegraphics[width=1\linewidth,height=0.6\linewidth]{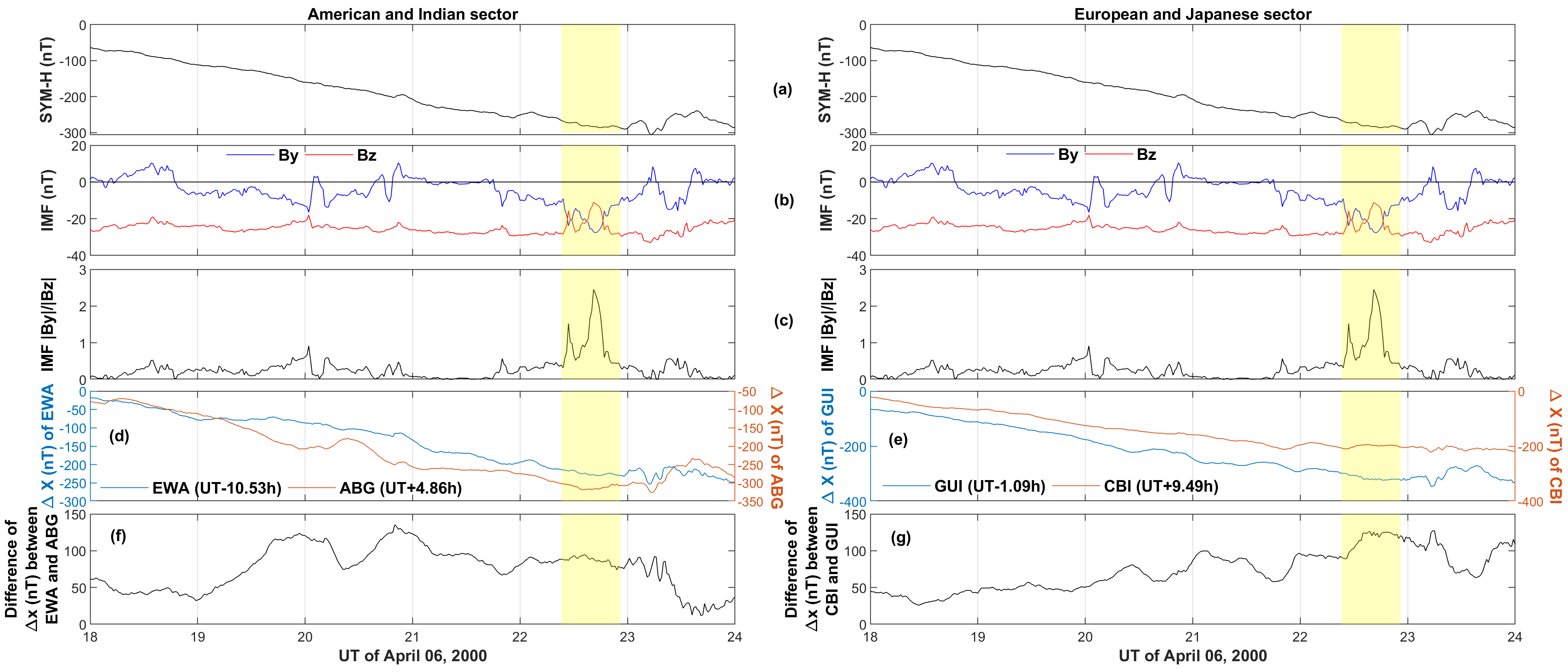}
\caption{Variations of (a) SYM-H (nT), (b) IMF $B_y$ (in blue) and IMF $B_z$ (in red), (c) ratio between the absolute values of IMF $B_y$ and IMF $B_z$, $\Delta$X (nT) variations between (d): EWA (in cyan) and ABG (in orange) and (e): CBI (in orange) and GUI (in cyan) and differences in the $\Delta$X variations (Diff($\Delta$X, nT) between (f): EWA and ABG and (g): CBI and GUI during 18:00 UT of April 06, 2000, and 00:00 UT of April 07, 2000. The yellow-shaded region of interest lies between 22:22 UT and 22:55 UT on April 06, 2000.} 
\label{sc01}
\end{figure}

\begin{figure}[ht]
\centering\includegraphics[width=0.75\linewidth,height=0.75\linewidth]{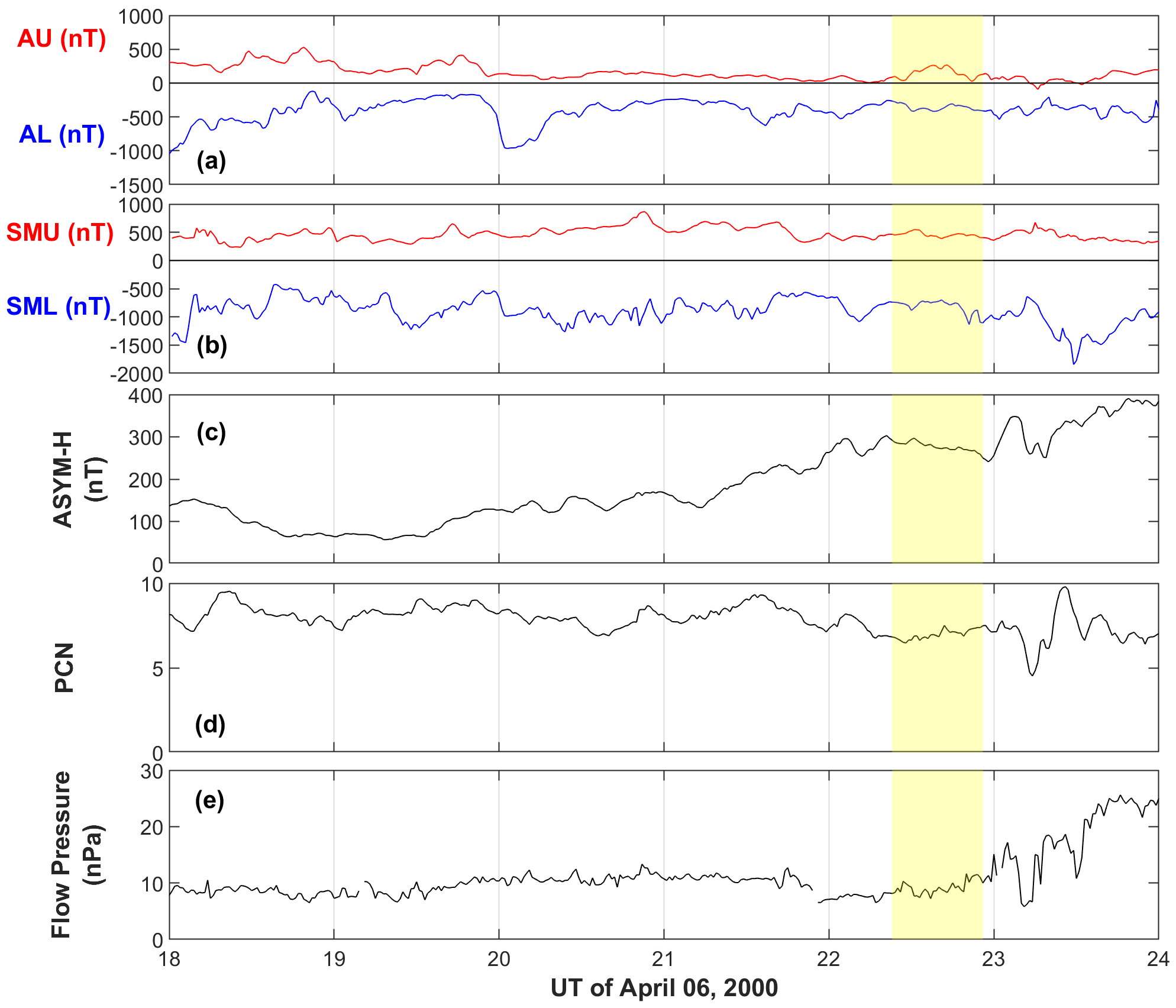}
\caption{Variations of (a) AU (nT, in red) and AL (nT, in blue), (b) SMU (nT, in red) and SML (nT, in blue), (c) ASYM-H (nT), (d) PCN and (e) solar wind flow pressure during 18:00 UT of April 06, 2000, and 00:00 UT of April 07, 2000. The yellow-shaded region of interest lies between 22:22 UT and 22:55 UT on April 06, 2000.} 
\label{sc02}
\end{figure}

A CME erupted near the western limb of the Sun on April 04, 2000, and hit the magnetosphere of Earth on April 06, 2000, causing one of the strongest geomagnetic storms of the solar cycle 23, being reported by several researchers \cite{sc:6,sc:7,sc:8,sc:9,sc:10,sc:11}. Figures \ref{sc01} and \ref{sc02} show the interplanetary and geomagnetic variations during the period from 18:00 UT of April 06, 2000, to 24:00 UT (or 00:00 UT) of April 07, 2000. The selected period of observation is in the main phase of the strong geomagnetic storm as can be observed in panel \ref{sc01}(a). Figure \ref{sc01}(b) shows the corresponding IMF $B_y$ (in blue) variation and IMF $B_z$ (in red) variation which can be observed to be completely southward throughout the period. Figure \ref{sc01}(c) shows the ratio between IMF $|B_y|$ and IMF $|B_z|$. Figure \ref{sc01}(d) shows the $\Delta$X component of the magnetic field as measured by the magnetometers over nearly antipodal locations: EWA (in blue) and ABG (in red) while Figure \ref{sc01}(e) shows the same over another set of nearly antipodal locations: GUI (in blue) and CBI (in red). The Local Times (LT) of these locations with respect to the Universal Time (UT) are presented in Table \ref{sc00} as well as marked in Figure panels \ref{sc01}(d) and (e). Figure \ref{sc01}(f) shows the difference of $\Delta$X between the locations EWA and ABG while Figure \ref{sc01}(g) shows the same between the locations CBI and GUI. This difference has been calculated by subtracting the $\Delta$X variation over the location at the nightside from the $\Delta$X variation over the location at the dayside. For the same period, Figure \ref{sc02}: panels (a) to (d) show the Auroral Upper (AU, in red) and the Auroral Lower (AL, in blue), the ASYM-H, the Polar Cap index (PCN) and the solar wind flow pressure respectively. 

The region of interest has been shaded in yellow in both Figures \ref{sc01} and \ref{sc02}. The duration of this shaded region is from 22:22 UT to 22:55 UT on April 06, 2000. The motivation for selecting this region is due to the fact that it is the only region where the magnitude of IMF $B_y$ can be observed to be significantly higher than that of IMF $B_z$. During this interval, the SYM-H (Figure \ref{sc01}(a)) value ranged from  -266 nT to -287 nT. Corresponding IMF $B_y$ values were -11.08 nT and -10.05 nT (both negative)  while the IMF $B_z$ values (both southward) were -27.94 nT and -29.08 nT (Figure \ref{sc01}(b)). 
To understand the relative roles of the two IMF components, the ratio of the absolute values of IMF $B_y$ to IMF $B_z$ has been calculated and plotted (Figure \ref{sc01}(c)). During this particular interval, both the IMF components were negative thus suggesting calculation of absolute values of the ratio and the calculation of a simple ratio between these two IMF components would bring out the same result. 
The two peaks observed in this panel (c) occur at 22:26 UT and 22:40 UT with values of 1.52 and 2.45 which suggests that during this interval, although both were negative, the IMF $B_y$ magnitude had been 1.5-2.5 times that of the magnitude of IMF $B_z$. The $\Delta$X variations between the pair of nearly antipodal locations (EWA and ABG) and (CBI and GUI) can be observed to be correlated, which is in agreement with that reported in \cite{sc:22} for nearly antipodal locations over these latitudinal regions due to the effect of IMF $B_y$ under southward IMF $B_z$. It is to be noted that during this interval of time, the LTs of these four stations were as follows: EWA (11:50-12:23 of April 06, 2000), ABG (03:14-03:47 of April 07, 2000), CBI (07:52-08:15 of April 07, 2000) and GUI (21:17-21:50 of April 06, 2000) thus pointing to the fact that EWA and CBI were in dayside while ABG and GUI were in the nightside. Taking reference to these LTs, the difference between the $\Delta$X (dayside-nightside) variations: Diff($\Delta$X), between the two pairs of nearly antipodal stations are plotted in the panels (f) and (g) of the same figure where it can be observed that during this interval, the difference is almost constant (around 90 nT) for the EWA and ABG pair while an increase from about 90 nT to 125 nT is observed for the CBI and GUI pair of stations. Further, it is important to note that during the first peak of the ratio between IMF $|B_y|$ and IMF $|B_z|$, which should have a value of 1.52, the Diff($\Delta$X) between EWA and ABG was 86.8 nT and that between CBI and GUI was 96.7 nT, however, when the second peak had occurred (value of the ratio: 2.45), the Diff($\Delta$X) between CBI and GUI increased to 125.5 nT although the same between EWA and ABG was 89.0 nT. Thus an asymmetry in the magnetic field variations between these two pairs of nearly antipodal locations can be observed due to the combined effects of IMF $B_z$ and IMF $B_y$. 

Figure \ref{sc02} has been presented to observe for any substorm and/or pressure-related effect during the period of interest. Based on the criteria defined by \cite{sc:54,sc:55}, using SML index (Figure \ref{sc02}(b)) we can infer occurrence of a substorm around 22:40 UT. However, we do not see any magnetic bay-like disturbances \cite{sc:56,sc:57} at this time in the $\Delta$X variation (Figure \ref{sc01}(d)) over the nightside stations. These suggest that the $\Delta$X variations at this time over these stations are not significantly affected by substorms. In addition, we also do not observe any sharp changes in the ram-pressure (Figure \ref{sc02}(e)) at this time which will cause transient electric field perturbations and associated changes in the $\Delta$X variations.   

\begin{figure}[ht]
\centering
\includegraphics[width=0.6\linewidth,height=1.2\linewidth]{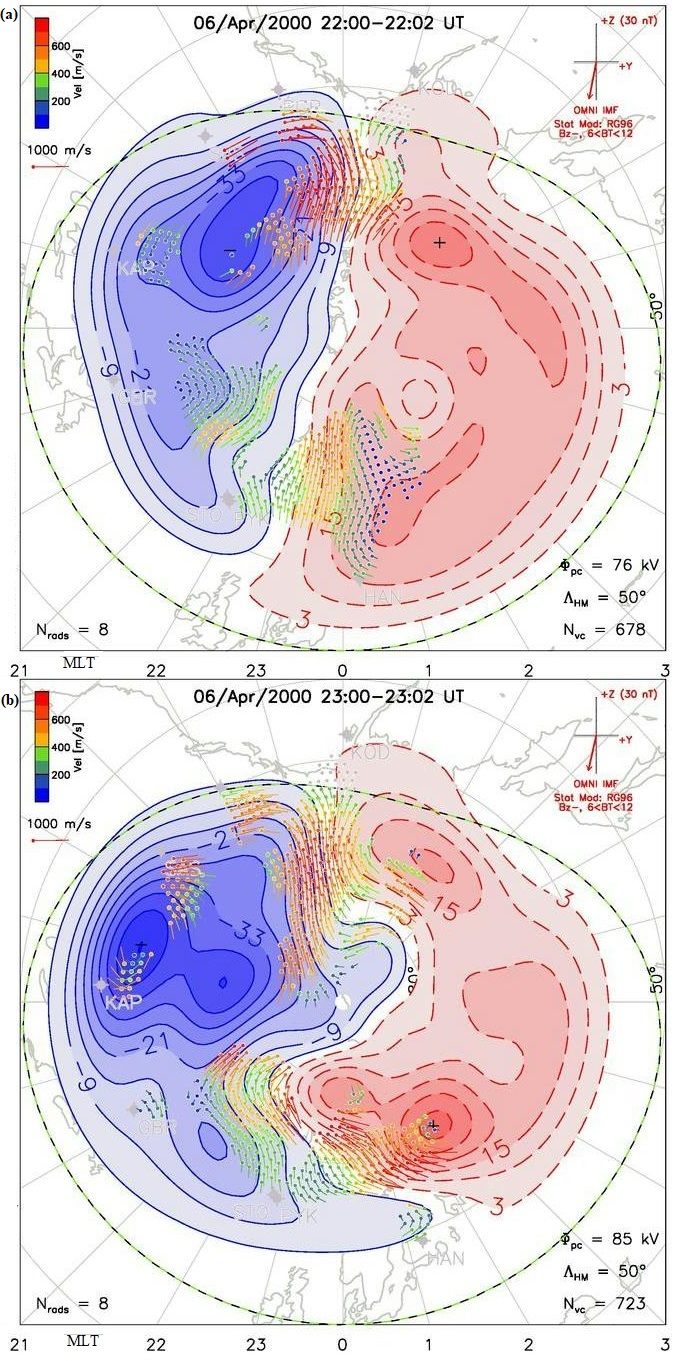}
\caption{SuperDARN ionospheric convection maps along with DP2 contours during (a) 22:00–22:02 UT and (b) 23:00–23:02 UT of April 06, 2000.}
\label{sc03}
\end{figure}

To validate the fact that there had been a relative role played by both IMF $B_z$ and IMF $B_y$ during this particular interval that had caused such asymmetry in the $\Delta$X variations, Figure \ref{sc03} has been shown which consists of the SuperDARN ionospheric convection maps at time stamps: (a) 22:00-22:02 UT and (b) 23:00-23:02 UT of April 06, 2000. Anti-clockwise rotation, distortions of both the dawn (in red) and the dusk (in blue) cells, as well as the electrodynamic divider (due to southward IMF $B_z$ and dawnward IMF $B_y$), can be observed on the map of Figure \ref{sc03}(b) with respect to the map of Figure \ref{sc03}(a). Further, in \ref{sc03}(b) one can observe distinctly enhanced throat flows in between the two cells (see \citeA{sc:22} and references therein).

It is important to note that the SuperDARN maps available during the study period are of low (1 hour) resolution and further, these maps give the velocity vectors which are not suitable for the present study which is focused on the observations of the $\Delta$X variations and associated equivalent currents. Therefore, for performing global studies of the terrestrial MI coupling and the associated signatures of the global DP2 currents related to the interplay effect between the two components of IMF, one has to look into the two-dimensional maps of equivalent currents that are derived using the global magnetometer data. These maps, which represent the global perturbation magnetic field, are obtained by performing a spherical cap harmonic expansion of the SuperMAG magnetic field data. The procedure to obtain these two-dimensional maps are explained in detail in \cite{sc:52}. 

\begin{figure}[ht]
\centering
\includegraphics[width=0.495\linewidth,height=0.495\linewidth]{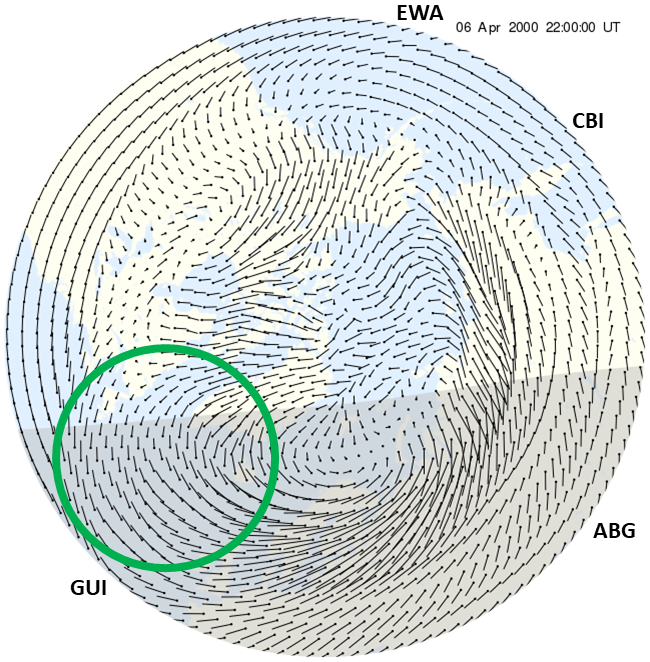}
\includegraphics[width=0.495\linewidth,height=0.495\linewidth]{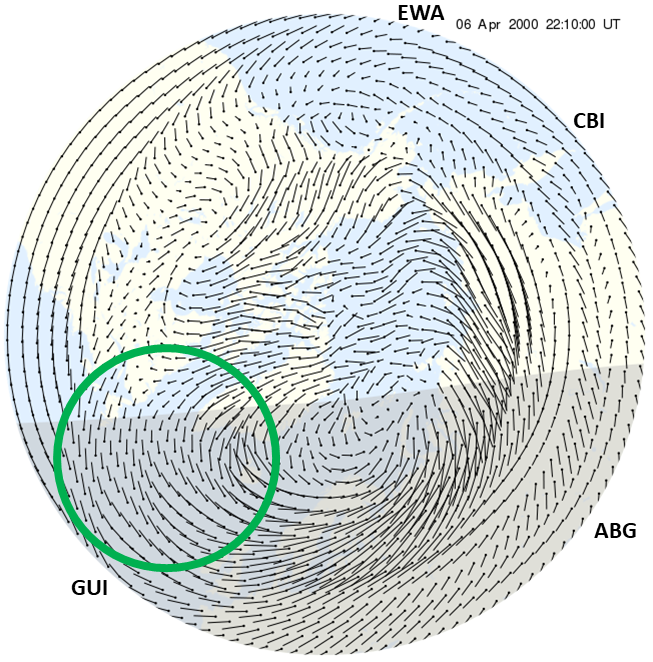}
\includegraphics[width=0.495\linewidth,height=0.495\linewidth]{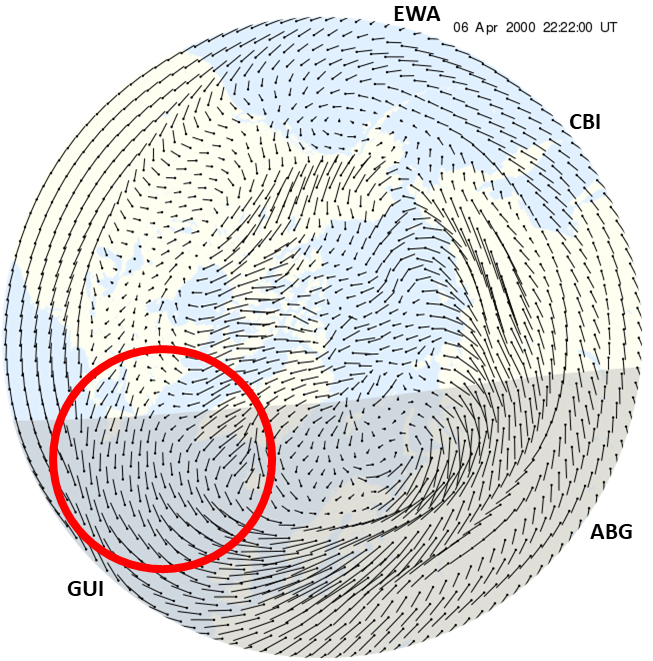}
\includegraphics[width=0.495\linewidth,height=0.495\linewidth]{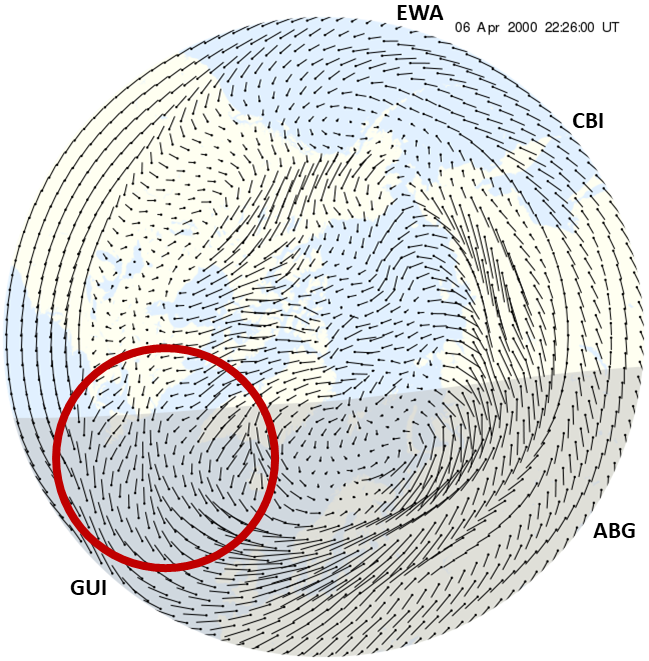}
\caption{The two-dimensional maps of equivalent current as obtained from the global magnetometer data during 22:00 UT, 22:10 UT, 22:22 UT, and 22:26 UT of April 06, 2000. Anomalous orientations of the current vectors and associated throat flow regions are designated in light and dark shades of red circles, while the same for usual orientations are designated in green circles.}
\label{sc04}
\end{figure}

\begin{figure}[ht]
\centering
\includegraphics[width=0.495\linewidth,height=0.495\linewidth]{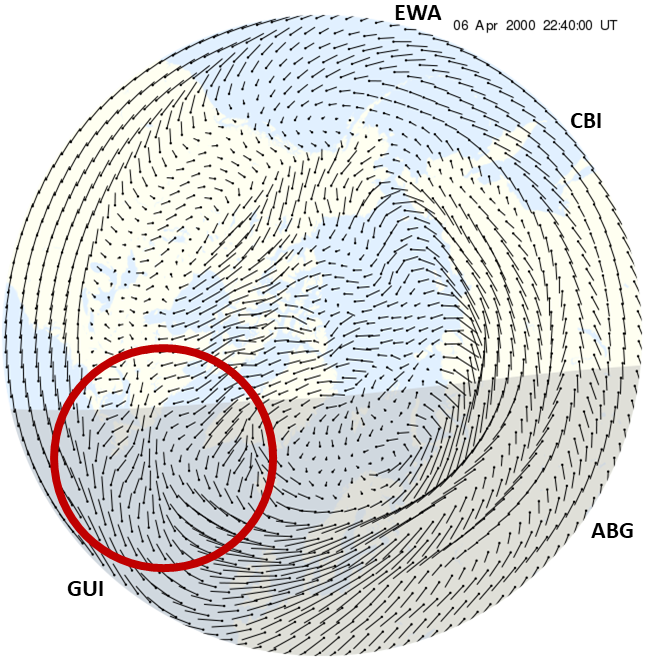}
\includegraphics[width=0.495\linewidth,height=0.495\linewidth]{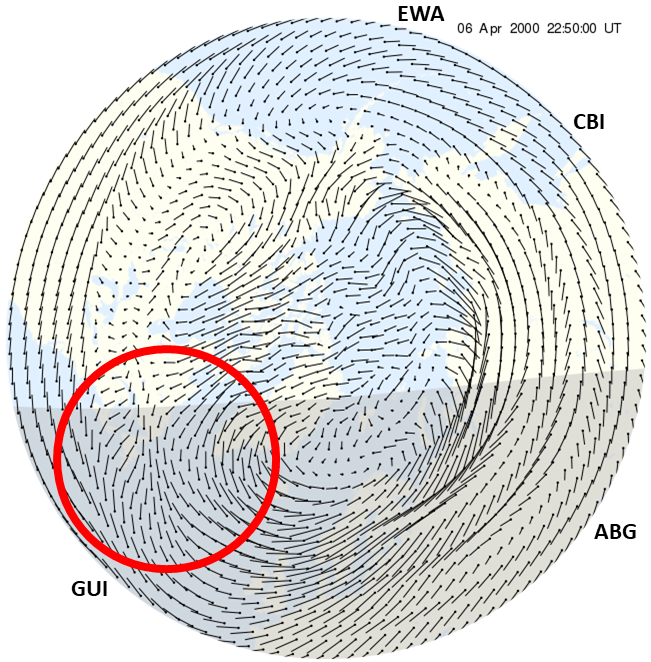}
\includegraphics[width=0.495\linewidth,height=0.495\linewidth]{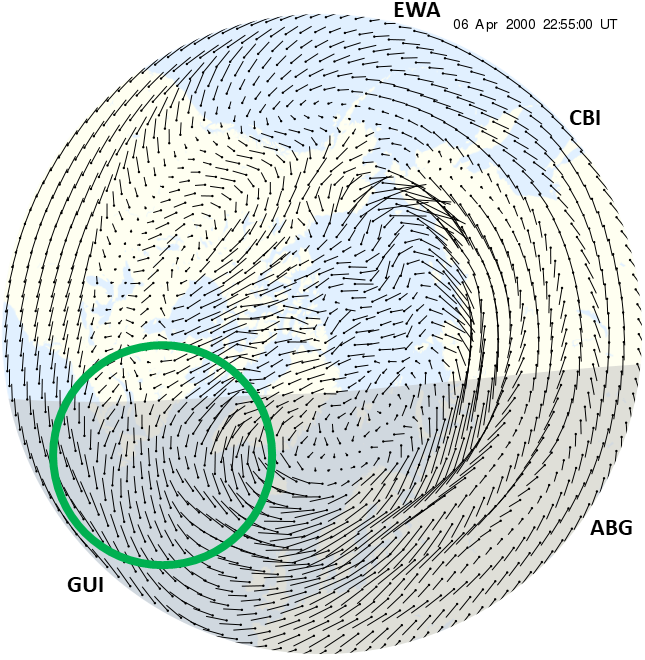}
\includegraphics[width=0.495\linewidth,height=0.495\linewidth]{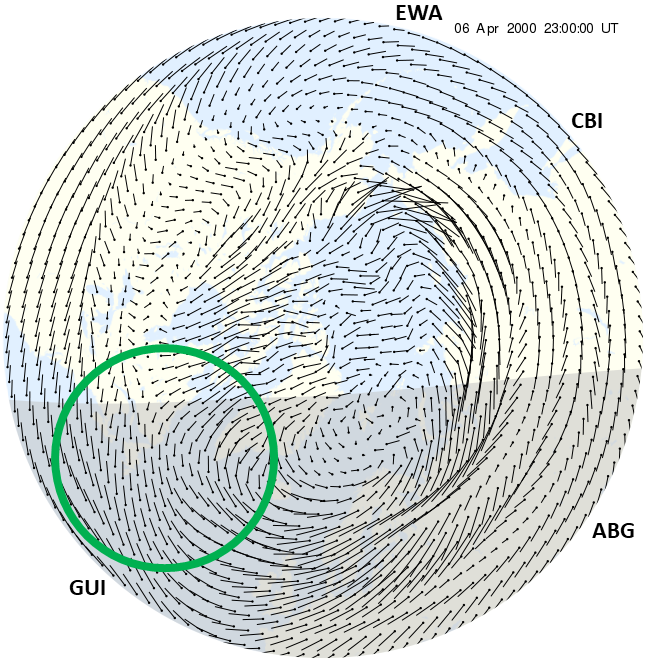}
\caption{Same as Figure \ref{sc04} but during 22:40 UT, 22:50 UT, 22:55 UT, and 23:00 UT of April 06, 2000.}
\label{sc05}
\end{figure}

The maps of equivalent currents used for this study are shown in Figures \ref{sc04} and \ref{sc05}, during intervals between 22:00 UT to 23:00 UT of April 06, 2000. In all the panels, the locations of the four stations (following their MLT distributions) have been marked. The grey-shaded regions in all the maps mark the nightside. The colored circles show the region where changes are observed. The maps in the panels at the time instant of 22:00 UT and 22:10 UT in Figure \ref{sc04} show a green circle marked inside which the usual orientation of the current vectors can be observed. Coming to the time instant of 22:22 UT, a slight distortion (inside the light red circle) in these current vectors can be observed. Further, at the time instant of 22:26 UT, there is a clear observation of a distortion feature in the current vectors that become radially outward (inside the dark red circle) over the station GUI which is different compared to the current vectors over stations EWA, ABG and CBI. During 22:40 UT (first map of Figure \ref{sc05} and inside the dark red circular region) the distortions of the current vectors over GUI increase with respect to that observed during the previous time instant (22:26 UT map in Figure \ref{sc04}). Interestingly, at 22:26 UT and 22:40 UT, there were occurrences of the first and the second (higher in magnitude) peaks in the ratio of IMF $B_y$ and IMF $B_z$. The map at the time instant of 22:50 UT shows a continuation of these distortions in the current vectors. The vectors then get restored to their usual orientations in the maps during the time instances 22:55 UT and 23:00 UT. Further explanations of these results will be discussed in the following section.

\clearpage
\section{Discussion}

Before discussing the probable mechanism to have caused the observed asymmetry in the present study, it is important to rule out the effects of ring current asymmetry during the main phase of the geomagnetic storm as well as the effects of the disturbance dynamo electric fields on the magnetic field variations.

There could be effects of asymmetry of the ring current on the magnetic field perturbations observed during the main phase of geomagnetic storms over the low-latitudes as shown by \cite{sc:51}, however, the asymmetry would occur near the dawn and the dusk sector. The distribution of the stations selected is away from the dawn-dusk sector. Furthermore, during the geomagnetic storm main phase, \cite{sc:43} have shown that there would be changes (away from the dawn-dusk sector in the MLT distribution of the ring current as observed at the ground as a result of the influence of IMF $B_y$ under southward IMF $B_z$, however under the condition that IMF $B_y$ has to be of the same orientation for longer timescales (at least 12 h), which is not the case in the present study. It can also be observed from Figure \ref{sc02}(c) that there was a steady decrease of the ASYM-H during the period of interest, which supports the fact that the observations are due to changes in the ionospheric currents and not due to the intensification of partial ring current.

Additionally, as the effects of the disturbance dynamo electric fields are longitude dependent, there would be some contribution of the same in the nighttime stations' (one in the post-sunset sector and the other in the post-midnight sector) $\Delta$X variations. However, we believe that these would not be able to produce the changes observed in the present study in the differences of the $\Delta$X variations of one pair with respect to the other pair of nearly antipodal locations as the event under consideration occurs for a short period of time and that too, rather abruptly. Therefore, it is unlikely that disturbance dynamo produces such transient and sharp changes in the $\Delta$X variations reported in the present case.

In the context of the evolution of the DP2 cells, it is well-known \cite{sc:5,sc:53,sc:22} that under southward $B_z$, the dawn and the dusk cells take the circular/orange and the crescent/banana-like shapes having a distinct anti-sunward throat flows in between these two cells. When the IMF $B_y$ is negative (or turns dawnward), there is an overall anti-clockwise rotation of the electrodynamical divider with respect to the noon-midnight characteristic orientation when IMF $B_y$ = 0, further resulting in a clockwise rotation of the throat flows. In the present study, the occurrence of enhanced throat flows can be observed in Figure \ref{sc03}. However, the SuperDARN maps during this period are of low cadence and they only represent the velocity vectors which are not appropriate for studying the variations in the global current system under the effects of the relative roles of IMF $B_z$ and $B_y$. As a result, two-dimensional maps are used in the present study. Interestingly, on all the maps in Figures \ref{sc04}  and \ref{sc05}, the orientation of the current vectors near EWA, CBI, and ABG are nearly similar whereas, over GUI, the current vectors become distorted/radially outward during the period of interest, especially more distorted when there was the observation of sharp peaks in the ratio of IMF $B_y$ to IMF $B_z$. This suggests that as the station GUI was under the region where there were radially outward currents during these intervals, differences in the corresponding ground magnetic field variation were observed which resulted in an asymmetry when observing Diff($\Delta$X) in the pair of stations: CBI and GUI with respect to the pair of stations: EWA and ABG.

\begin{figure}[ht]
\centering\includegraphics[width=0.75\linewidth,height=0.85\linewidth]{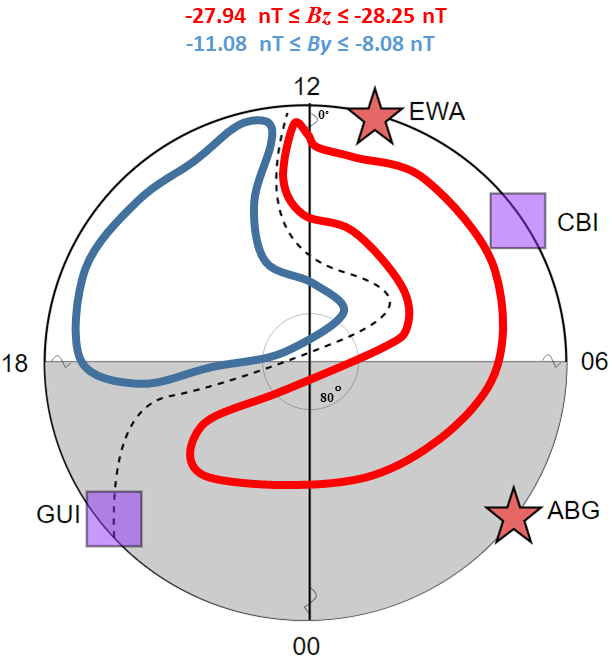}
\caption{A schematic of the DP2 equipotential contour under southward IMF Bz and negative IMF By condition during the region of interest. The shaded (grey) hemisphere represents the nightside. The solid black line represents the midnight-noon meridian while the dashed black line represents the electrodynamical divider between the dawn (red) and dusk (blue) cells. The nearly antipodal station pairs are marked in violet squares for CBI and GUI and red stars for EWA and ABG.}
\label{sc06}
\end{figure}

Finally, to understand the probable orientation of the electrodynamic divider, the shapes of the cells and the location of the stations, Figure \ref{sc06} shows the schematic of the DP2 equipotential contour for the conditions under southward IMF $B_z$ and negative IMF $B_y$ (range of values highlighted on top of in the figure for ready reference) during the region of interest (22:22 UT to 22:55 UT of April 06, 2022). The grey-shaded hemisphere here represents nighttime, the solid black line represents the midnight-noon meridian while the dashed line in black indicates how the electrodynamic divider had rotated during this particular period. The most important point of this schematic is the locations of the four stations with respect to the electrodynamic divider, wherein although both the pairs are at nearly antipodal locations (CBI and GUI: marked as violet squares and EWA and ABG: marked as red stars and placed according to their MLT), asymmetry in the Diff($\Delta$X) variations of CBI and GUI is observed with respect to EWA and ABG since GUI is the only station falling in the throat flow region. Therefore, an important observation comes out from this schematic which suggests that antipodal locations (day-night separator) are not a sufficient factor but the electrodynamical day-night or the electrodynamic divider and associated throat flow regions, are the factor that would decide how the changes in IMF $B_y$ and IMF $B_z$ and their relative roles that had distorted as well as rotated the DP2 cells in such a manner to present asymmetry between the differences in the pairs of nearly antipodal stations over these regions throughout the globe. It is crucial to note that the role of IMF $B_y$ over these locations comes as a secondary effect and sparse studies of such cases can be explained by the fact that one has to rule out the other drivers like substorm, solar wind ram pressure, ring current asymmetry, disturbance dynamo, etc, thus making statistical studies of such to be highly non-trivial. However, identifying more such cases, as in the present study, would be helpful for the community to perform a comprehensive statistical study of the global MI coupling in the near future. 

\clearpage
\section{Summary}

The present work was directed to understand whether there would be similarities or asymmetries in the magnetic field variations of any given pairs of antipodal stations under the relative roles of IMF $B_z$ and IMF $B_y$. The results showed that it is the changes in the spatial distribution of the equivalent currents and the subsequent distortions in the DP2 cells in addition to the rotation of the electrodynamical divider, due to the effects of the interplay between the two IMF components, that determines the asymmetries observed in the $\Delta$X variations in one pair of nearly antipodal stations with respect to the other pair. This study would be pertinent for developing a reliable space weather forecast system as it highlights the importance of the interplay between the two IMF components in determining the ionospheric impact over low latitudes during strong geomagnetic conditions.

\acknowledgments

The authors acknowledge the use of magnetic data, the two-dimensional maps of equivalent currents and the SMU/L indices openly available from the SuperMAG network website (http://supermag.jhuapl.edu/). The authors further thank the PIs of the magnetic observatories and the national institutes that support the observatories. The authors also acknowledge the use of SuperDARN data openly available from the Virginia Tech SuperDARN website (http://vt.superdarn.org/).  SuperDARN is a collection of radars funded by national scientific funding agencies in Australia, Canada, China, France, Italy, Japan, Norway, South Africa, the United Kingdom, and the United States of America. Further acknowledgments go to the NASA OMNIWEB Data Explorer website (https://omniweb.gsfc. nasa.gov/form/omni\_min.html) for the high resolution (1 minute) IMF $B_y$, IMF $B_z$, AU, AL, SYM-H, ASYM-H, PCN and solar wind flow pressure data. The authors thank both reviewers for their valuable comments and suggestions that have enhanced the quality of the manuscript. This research work is supported by the Department of Space, Government of India.

\bibliography{PDF-P1.bib}

\end{document}